\documentclass[amssymb,aps,twocolumn,prb]{revtex4}

\usepackage{epsfig}
\usepackage{graphicx}
\usepackage{dcolumn}
\usepackage{bm}

\input{tcilatex}

\begin{document}
\title{Screening ineffectiveness and THz emission at bare LO phonon frequencies}
\author{F. M. Souza and J. C. Egues}
\email{egues@if.sc.usp.br} \affiliation{Departamento de
F\'{\i}sica e Inform\'{a}tica, Instituto de F\'{\i}sica de S\~{a}o
Carlos, \\Universidade de S\~{a}o Paulo, 13560-970 S\~{a}o Carlos,
S\~{a}o Paulo, Brazil.}

\begin{abstract}
Within a hydrodynamic approach we investigate the dynamics of an
inhomogeneous electron-hole gas coupled to phonons in Te and the
corresponding THz emission. We find that the {\em longitudinal} {\em %
inhomogeneity} of the photogenerated electron-hole gas -- due to short
absorption lengths in Te -- gives rise to {\em screening ineffectiveness}
for non-zero wave-vector modes. This allows for THz dynamics and emission at
the bare LO phonon frequency $\nu _{LO}$ even at high carrier densities.
Screening ineffectiveness thus provides an appealingly simple explanation
for the existence of bare modes at $\nu _{LO}$ in longitudinally
inhomogeneous systems such as Te; no lateral inhomogeneity of the excitation
spot is needed here.
\end{abstract}

\date{\today}
\maketitle

Terahertz-range dynamics of non-equilibrium carriers interacting
with phonons is a very active field of research in
semiconductors.\cite {dekorsy-review} Pump-and-probe time-resolved
spectroscopies with ultrashort laser pulses have played a crucial
role in unveiling many interesting features in these complex
systems. The appearance of time-dependent electric fields
following the pump excitation -- due to screening of
depletion-layer fields, charge-separation fields, lattice
polarization fields, etc. -- gives rise to electro-optic induced
reflectivity/transmittance changes and terahertz emission due to
dipole radiation. Experimentalists probe the system in essentially
two ways: by measuring (i) the reflectivity/transmittance changes
of a probe pulse and/or (ii) the emitted radiation. These
canonical experimental geometries yield, in principle,
complementary information on the system characteristic
frequencies.

In {\em n}-doped GaAs, for instance, the usual two branches $L_{+}$ and $%
L_{-}$ describing the {\em long wavelength} plasmon-phonon coupled
modes as a function of the carrier density are easily mapped out
from electro-optic modulation experiments.\cite{studies} As
pointed out in Ref. \onlinecite{yee}, the $L_{+}$ mode is usually
interpreted as a {\em bare} LO phonon mode $\nu _{LO}$ at low
carrier densities, while the $L_{-}$ mode is referred to as a {\em
screened} LO\ phonon mode $\nu _{TO}$ at high carrier densities.
Interestingly enough, beating of these TO\ and LO modes is
observed in the experiments. This is due to the simultaneous
excitation of these modes due to the {\em lateral} inhomogeneity
of the pump pulse which generates regions of high and low carrier
densities -- sampled simultaneously by the probe pulse. In
contrast, terahertz emission spectroscopy in GaAs provides
information only on the $L_{-}$ branch at low densities; $L_{+}$
and LO phonon frequencies ($>8$ THz) are not accessible with the
current spectral range ($\lesssim 7$ THz) of the state-of-the-art
antennas.\cite{heynman}

In Tellurium on the other hand, both reflectivity and THz emission
measurements can probe coupled-mode frequencies.\ These
experiments are performed in both the c$_{\parallel }$ and
c$_{\perp }$ geometry. These denote the either parallel or
perpendicular orientation of the c axis of the hexagonal lattice
of Te with respect to the excitation surface. Oscillations at bare
$\nu _{LO}$ are seen in both experiments for the c$_{\parallel }$
geometries.\cite{dekorsy0} In the c$_{\perp }$ geometry only THz
emission is possible; a peak at $\nu _{LO}$ is also
evident.\cite{dekorsy},\cite{tani} Similarly to GaAs, the
existence of these peaks in the Te response can in principle be
explained via the {\em lateral}{\bf \ }inhomogeneity of the
excitation spot. There is, however, a more appealing explanation
we propose in this work.

In this letter we theoretically investigate the coupled dynamics of
electron-hole--phonon modes in\ Te and the corresponding THz emission. Our
semiclassical description incorporates phonons and its coupling to carriers
via polarization charges in Poisson's equation. Our main finding is the
observation that the intrinsic {\em longitudinal} inhomogeneity of the
photogenerated electron-hole plasma -- due to the short absorption lengths
in Te -- leads to {\em screening ineffectiveness} for non-zero wave vectors.
Screening ineffectiveness allows for dynamics and terahertz emission at bare
LO phonon frequencies -- even at {\em high} carrier densities -- thus
providing a possible alternate/complementary explanation for the observed
bare LO modes in the THz response of Te. We suggest a simple experiment to
tell apart lateral- ($q=0$) and longitudinal-inhomogeneity ($q\neq 0$)
contributions. A dielectric function analysis (Debye H\"{u}ckel and RPA)
corroborates our numerical findings.

{\em Ingredients.} Any reasonable attempt to describe pump-and-probe THz
emission experiments has to necessarily address the coupled-field dynamics
of electrons, holes, and phonons. Our semiclassical approach combines (i)
moments of the Boltzmann equation describing charge and current densities,
(ii) \ a phenomenological driven-harmonic-oscillator equation simulating LO
phonons, and (iii) Poisson's equation providing electron- and hole-phonon
coupling via polarization charges.

{\em Model. }Our one-dimensional\cite{onedimensional} model allows
motion only perpendicular to the relevant semiconductor surface
being excited. We
consider carriers with isotropic masses $m_{e}^{\ast }=0.06m_{0}$ and $%
m_{h}^{\ast }=0.114m_{0}$ (and charges $q_{e}$ and $q_{h}$) in parabolic
bands.\cite{alex} We assume the electrons and holes are thermalized to distinct {\em %
constant} temperatures $T_{e}$ and $T_{h}$,
respectively;\cite{temperature} these are, however, different from
the lattice temperature. By considering the semiclassical
Boltzmann equation for the distribution function within the usual
relaxation-time approximation, we derive equations for both
particle and particle current densities of electrons ({\em e}) and holes (%
{\em h}), $n_{e(h)}(x,t)$ and $J_{e(h)}(x,t)=n_{e(h)}(x,t)v_{e(h)}(x,t)$,
respectively; $v_{e(h)}(x,t)$ denotes the electron (hole) mean velocity.
These are
\begin{equation}
\frac{\partial n_{e(h)}(x,t)}{\partial t}+\frac{\partial J_{e(h)}(x,t)}{%
\partial x}=G(x,t),  \label{eq1}
\end{equation}
and
\begin{eqnarray}
\frac{\partial J_{e(h)}(x,t)}{\partial t} &=&-\frac{\partial }{\partial x}%
\frac{J_{e(h)}^{2}(x,t)}{n_{e(h)}(x,t)}+\frac{q_{e(h)}}{m_{e(h)}^{\ast }}%
n_{e(h)}(x,t)E(x,t)-  \nonumber \\
&&\frac{k_{B}T_{e(h)}}{m_{e(h)}^{\ast }}\frac{\partial }{\partial x}%
n_{e(h)}(x,t)-\frac{J_{e(h)}(x,t)}{\tau _{e(h)}},  \label{eq2}
\end{eqnarray}
where $G(x,t)$ is the generation term (laser). The electric field $E(x,t)$
in Eq. (\ref{eq2}) is calculated via Poisson's equation
\begin{equation}
\frac{\partial E}{\partial x}=\frac{1}{\varepsilon _{0}\epsilon _{\infty }}%
(q_{h}n_{h}+q_{e}n_{e})-\frac{\omega _{TO}\sqrt{\varepsilon _{0}(\epsilon
_{0}-\epsilon _{\infty })}}{\varepsilon _{0}\epsilon _{\infty }}\frac{%
\partial w}{\partial x},  \label{eq3}
\end{equation}
which accounts for both electron and hole charge densities and lattice
polarization charges $\sim \partial w(x,t)/\partial x$. We {\em assume} the
vibrational coordinate $w(x,t)$ in Poisson's equation obeys
\begin{equation}
\frac{\partial ^{2}w}{\partial t^{2}}+\frac{1}{\tau _{ph}}\frac{\partial w}{%
\partial t}+\omega _{TO}^{2}w=\omega _{TO}\sqrt{\varepsilon _{0}(\epsilon
_{0}-\epsilon _{\infty })}E,  \label{eq4}
\end{equation}
which is driven by the {\em total} electric field due to electron-hole
charge separation and phonon polarization charges. Note that Eq. (\ref{eq4})
describes longitudinal oscillations. We account for damping of electrons,
holes and phonons via proper relaxation times [$\tau _{e}$, $\tau _{h}$ and $%
\tau _{ph}$, respectively].

Equation (\ref{eq1}) is the continuity equation for electrons \ or holes;
laser pumping of carriers is accounted for by the generation term (gaussian
in time and exponentially attenuated in $x$). Equation (\ref{eq2}) states
momentum conservation [the momentum density is $%
p_{e(h)}(x,t)=n_{e(h)}(x,t)m_{e(h)}^{\ast }v_{e(h)}(x,t)$]. Implicit in (\ref
{eq2}) is the assumption of thermalized carriers with average energies $%
\langle u_{x}\rangle _{e(h)}=\frac{1}{2}k_{B}T_{e(h)}+\frac{1}{2}%
m_{e(h)}^{\ast }v_{e(h)}^{2}(x,t)$. With this assumption we are
essentially truncating the sequence of equations for higher-order
moments. By coupling Eqs. (\ref{eq3}) and (\ref{eq4}), we extend
the phenomenological description of Born and Huang\cite{huang} in
that we allow the internal electric field to drive lattice
vibrations and vice versa. Our description here generalizes the
previous investigations of Kuznetsov and Stanton\cite{alex} who
treat homogeneous systems and introduce averaging over the spot
profile and
Dekorsy {\em et al.}\cite{dekorsy} who use a drift-diffusion type model.{\em %
\ }

{\em Numerical solution.} The set of six coupled differential
equations discussed above is tricky to solve numerically. By
carefully using standard explicit schemes for the propagation of
partial differential equations,\cite{nrc} we are able to obtain $n_{e}(x,t)$, $J_{e}(x,t)$, $n_{h}(x,t)$, $%
J_{h}(x,t)$, $E(x,t)$, and $\ w(x,t)$ in a wide range of system parameters.
As we discuss next, both the radiated signal and the reflectivity response
are related to $E(x,t)$.

{\em Emitted THz radiation.} The longitudinal charge oscillations in the
system are essentially electric dipoles and thus emit electromagnetic waves.
Within the dipole approximation, the emitted signal is proportional to the
second time derivative of the internal electric field. We define the
radiated field $E_{rad}$ by
\begin{equation}
E_{rad}\propto \frac{\partial ^{2}}{\partial t^{2}}\int\limits_{0}^{\alpha
^{-1}}e^{-\alpha x}E(x,t)dx,  \label{eq5}
\end{equation}
which accounts for oscillating dipoles, weighed by the absorption
profile, from within only the relevant absorption length $\alpha
^{-1}$. Note that the radiated field $E_{rad}$ is {\em p}
polarized.

Figures 1(a) and 1(b) show the calculated {\em emitted} signals and their
Fourier transform, respectively, for several excitation densities {\em n}$%
_{exc}$, at fixed electron and hole temperatures, $T_{e}=2000$ K and $%
T_{h}=1050$ K [see also inset in Fig. 1(c)]. The time-dependent
signals 1(a) show an initial rise and a dip followed by small
oscillations [similar to the experimental detected signal in the
inset]. As shown in 1(b), these small oscillations have
frequencies close to that of the LO\ phonon in$\;$Te, $\nu
_{LO}\sim 2.82$ THz ($c_{\perp }$ geometry); the broad-band peak
in 1(b) is due to electron-hole charge separation as discussed in
\onlinecite{dekorsy}.

The {\em detected} theoretical signals in 1(c) are convolutions of
the emitted signals in 1(b)\ and the antenna response [dotted line
in 1(b)]. The convoluted signals 1(c) are very similar to the
recent data in 1(d) from Ref. \onlinecite{dekorsy}. In particular
both theoretical and experimental\ spectra display a broad-band
peak followed by a dip at the screened LO\ frequency $\nu
_{TO}\sim 2.6$ THz and a subsequent peak at the {\em bare} LO
phonon frequency $\nu _{LO}\sim 2.82$ THz. Note that the signals
get stronger with increasing excitation densities [curves (1) --
(6)]. This intensity enhancement of the signals follows from the
amplitude increase of the internal fields with {\em n}$_{exc}.$
While the spectral features at the phonon frequencies hardly move
with increasing excitation densities, the broad-band features
slightly shift to higher frequencies. This shift is due to the
faster electron-hole charge dynamics at higher
densities.\cite{dekorsy} The relative intensities of the bare LO\
peak and that of the broad-band peak strongly depend on the
antenna response.

{\em Antenna response.} We assumed a simplistic antenna response with a high
sensitivity around 0.5 THz dropping essentially to zero above 3 THz [dotted
line in 1(b)]. This response weighs up the contribution of the broad-band
peak in contrast to that of the {\em bare }LO phonon. The
detected/convoluted signals 1(c) are then\ ``distorted'' thus displaying
weaker LO phonon peaks [cf. 1(b) and 1(c)].

{\em Normal mode analysis. }By linearizing our equations and neglecting
damping for simplicity, we obtain the system dielectric function
\begin{equation}
\epsilon (q,\omega )=\epsilon _{\infty }+\frac{\epsilon _{0}-\epsilon
_{\infty }}{1-\left( \frac{\omega }{\omega _{TO}}\right) ^{2}}%
-\sum\limits_{i=e,h}\frac{\omega _{pl,i}^{2}\epsilon _{\infty }}{\omega
^{2}-c_{i}^{2}q^{2}},  \label{eq6}
\end{equation}
where $\omega _{pl,i}=\sqrt{4\pi e^{2}n/\epsilon _{\infty }m_{i}^{\ast }}$, $%
c_{i}=\sqrt{k_{B}T_{i}/m_{i}^{\ast }}$, and $n$ is the carrier
densities. Equation (\ref{eq6}) contains both phonon and
electron-hole plasma contributions; these are the second and third
terms, respectively. The LO\ phonon contribution is
dispersionless. The plasmon contribution contains non-zero wave
vector modes which account for the {\em longitudinal
inhomogeneity} of the electron-hole gas; this term is equivalent
to the Debye-H\"{u}ckel screening (classical).\cite{rickayzen} The
zeros of the dielectric function define the normal modes of the
system:\ $\epsilon (q,\omega )=0\Rightarrow \omega (q)$.

Figure 2(a) displays the coupled carrier-phonon modes (solid
lines) of the system in the long wavelength limit ($q=0$) as a
function of the carrier density. The branches $L_{+}$ and $L_{-}$
arise from the anticrossing of the LO and plasmon modes (dashed
line) due to coupling.\cite{mooradian} Note that for increasing
{\em n }the $L_{-}$ branch\ tends to the screened LO phonon
frequency$\;\nu _{TO}.$ Note that screening is the strongest for
$q=0$ and high densities [Eq. (\ref{eq6})]. In the low-density
limit, in contrast, $L_{+}$ converges to the bare LO frequency
$\nu _{LO}$.

Figure 2(b) shows the wave-vector dependence of three particular $L_{-}$
frequencies corresponding to $n=2$, $4$, and $10\times 10^{17}$ cm$^{-3}$.
Here as $q$ increases the frequency of these three modes tend to the {\em %
bare} $\nu _{LO}$ value. This interesting feature is due to {\em
screening ineffectiveness} for non-zero wave vectors,\cite{cowley}
i.e., for large $q$ values screening by the electron-hole gas
becomes ineffective. Equation (\ref {eq6}) clearly shows the
electron-hole contribution to $\epsilon (q,\omega )$ gets
suppressed for increasing $q$. \ In addition, we find that a more
rigorous description of screening within the Random Phase
Approximation corroborates our semiclassical normal-mode analysis
[see dashed lines in Fig. 2(b)]. We consider only the static limit
for simplicity.\cite{rpa} In this limit and for the high
temperatures used, the RPA\ calculation is
straightforward.\cite{sato}

The mode analysis above makes it clear that the physics behind the
bare LO phonon mode in our numerical results (particularly at high
densities, Fig. 1) is that of screening ineffectiveness. We
believe this peak at $\nu _{LO}$ is an $L_{-}$ {\em unscreened}
$q\neq 0$ mode; not a low-density $L_{+}$ mode with $q=0$. Note
that this $q\neq 0$ phonon-like mode is not affected by the strong
damping of the electron-hole gas ($\tau _{e(h)}=10$ fs for
Te;\cite{dekorsy}) its damping constant is actually that of the
long-lived phonon modes (we {\em assume } $\tau _{ph}=4$
ps.\cite{alex})
Interestingly enough, the ``dip'' at $\nu _{TO}$ in\ Fig. 1 follows from \ $%
\epsilon (q,\omega )\rightarrow \infty $ as $\omega \rightarrow \omega _{TO}$
which leads to a suppression of the internal electric field. This
suppression is not complete in our simulation because of damping.

{\em Electro-optic modulation signal.}{\bf \ }The relevant\ ``probe signal''
in a pump-and-probe electro-optic modulation experiment is the quantity $%
S(t)\propto \frac{\partial }{\partial t}\int_{0}^{\alpha
^{-1}}e^{-\alpha x}[\gamma E(x,t)+\beta w(x,t)]dx$, where $\beta $
and $\gamma $ are (electro-optic) constants.\cite{alex} Figure 3
shows the Fourier spectrum of $S(t)$ for an {\em n}-doped GaAs
system\cite{Te-GaAs} with an {\it arbitrarily-chosen} small
absorption length $\alpha ^{-1}=10$ nm and no depletion layer
electric field. \cite{mechanism} In this case longitudinal
inhomogeneity is relevant. Similarly to the THz emission signal,
here we also clearly see a peak near the {\em bare} LO\ phonon
mode. This feature is again due to screening
ineffectiveness\cite{stolz} [see Fig. 2(b)]. Note that the
reflectivity spectrum is qualitatively different from the emission
one. No dip at the screened LO frequency $\nu _{TO}$ should be
expected in the reflectivity spectrum when both the phonon and
internal electric fields contribute to the response ($\left| \beta
/\gamma \right|
>1$ in GaAs). In this case, the response does not vanish when
$E(x,t)$ is suppressed. In our model, no lateral averaging over
the pump spot is needed in order to obtain peaks at $\nu _{TO}$
and $\nu _{LO}$.

{\em Lateral vs longitudinal inhomogeneity.} A pump-and-probe
reflectivity measurement with an excitation spot much larger than
the probe one\cite {gaas} (for Te a c$_{\parallel }$ geometry
should be used) can conclusively discern contributions from these
distinctive inhomogeneities. By focusing the probe right on the
high-density central region of the excitation spot, one should see
both bare and screened LO phonon peaks in the probe spectral
response. While the screened LO peak would represent the $L_{-}$ mode with $%
q=0$ [Fig. 2(a)], the bare LO peak should follow from screening
ineffectiveness due to $q\neq 0$ modes [Fig. 2(b)]. The presence
of this bare phonon mode at high densities would constitute a
unique signature for the relevance of the
longitudinal-inhomogeneity effect we propose in this paper.
Multiple quantum wells and superlattices are also suitable systems
to search for this effect;\cite{baumberg} the periodic modulations
along the growth direction give rise to longitudinal
inhomogeneity. Recent reflectivity experiments in\ GaAs
quantum-confined geometries\cite{yee} have found both screened and
bare LO modes; these experiments, however, use comparable pump and
probe spot sizes.

We have studied the coupled dynamics of electrons, holes and phonons within
a hydrodynamic approach. In longitudinally inhomogeneous systems such as Te,
we find that screening ineffectiveness at non-zero wave vectors allows for
oscillations at the {\em bare} LO\ phonon frequency in both reflectivity and
terahertz emission spectra. We believe this mechanism provides an
appealingly simple (alternate/complementary) explanation for the existence
of these bare LO modes even at high densities.

This work was supported by Funda\c{c}\~{a}o de Amparo \`{a}
Pesquisa do Estado de S\~{a}o Paulo (FAPESP). The authors
acknowledge discussions with L. Ioriatti, A. V. Kuznetsov, T.
Dekorsy and G. Cho. F.M.S. gratefully acknowledges a fellowship
from FAPESP/CAPES. The authors thank M. P. Hasselbeck for pointing
out Ref. [4] to them.

\begin{figure}[ht!]
\caption{Calculated THz signals for differing excitation
densities, in time (a) and frequency domains (b), (c), and (d).
The emitted spectra (b) display features at lower frequencies due
to electron-hole separation and at the characteristic phonon
frequencies $\protect\nu_{TO}$ and $\protect\nu_{LO}$. The
theoretical detected spectra (c) is obtained by convoluting the
emitted signals and the antenna response [dotted line in (b)]. The
calculated (c) and experimental (d) detected spectra are very
similar: for increasing excitation densities they both present a
strong broad band followed by a dip at the TO phonon frequency and
a weaker peak at the bare LO phonon frequency. Note also that the
signals get stronger for increasing excitation densities. Observe
that the relative intensities of the broad peak and that of the LO
peak are reversed after convolution [cf. (b) and (c)].
Surprisingly enough, a bare LO mode survives even at high densities $%
n\sim5\times10^{18}$ cm$^{-3}$; in our model this follows from {\it %
screening ineffectiveness} for non-zero wave vectors arising from
the intrinsic electron-hole {\it longitudinal inhomogeneity} in
Te. This effect does not qualitatively change with $T_e/T_h$ [see
inset in (c)]. The inset in (d) shows the experimental
time-dependent detected signal.} \label{fig1}
\end{figure}

\begin{figure}[ht!]
\caption{Electron-hole--phonon coupled modes. The $q=0$ modes (a)
show the usual two branches $L_-$ and $L_+$ due to anticrossing.
Here screening is operative thus making $\protect\nu_{LO}
\rightarrow \protect\nu_{TO}$ for high densities. Screening
ineffectiveness is clearly seen in (b) for non-zero wave vector
modes: as $q$ increases the $L_-$ branch approaches the bare
$\protect\nu_{LO}$ frequency. Both Debye-H\"uckel and RPA show
this trend, which is consistent with our numerical findings.}
\label{fig2}
\end{figure}

\begin{figure}[ht!]
\caption{Calculated reflectivity spectra for an {\em n}-doped
system with short absorption length. Illustratively we use GaAs
with an arbitrarily
small $\protect\alpha^{-1}$. Note that both $\protect\nu_{TO}$ and $\protect%
\nu_{TO}$ are present in the response. No averaging over the spot
profile is performed.} \label{fig3}
\end{figure}


\begin{references}
\bibitem{onedimensional}  The diameter of the excitation spot is much
greater than the absorption length.
\bibitem{temperature}  For typical carrier excess energies, we choose the
temperatures so that $T_{e}/T_{h}=m_{h}/m_{e}$. \cite{dekorsy}
\bibitem{Te-GaAs} We do not find plasmon modes in undoped systems (Te).
\bibitem{mechanism} Here we have in mind a clean UHV-cleaved surface (no
band bending due to Fermi-level pinning).
\end{references}
\end{document}